\documentclass[12pt,aps,prd,preprint]{revtex4}

\usepackage{epsfig}
\usepackage{amssymb}
\usepackage{amsmath}
\usepackage{amsfonts}
\usepackage{graphics}

\newcommand{\Refs}{Refs.}
\newcommand{\Ref}{Ref.}

\newcommand{\eq}{Eq.}

\newcommand{\fig}{Fig.}

\newcommand{\ket}[1]{\left| #1 \right>}
\newcommand{\dd}[2]{\frac{{\rm d}#1}{{\rm d}#2}}
\newcommand{\tdd}[2]{{\rm d}#1/{\rm d}#2}

\newcommand{\ie}{\emph{i.e.}}
\newcommand{\eg}{\emph{e.g.}}
\newcommand{\cf}{\emph{c.f.}}

\begin{document}

\title{Prospects for cosmic neutrino detection in tritium experiments in the case of hierarchical neutrino masses}

\author{Mattias Blennow}
\email[]{blennow@mppmu.mpg.de}
\affiliation{Max-Planck-Institut f\"ur Physik (Werner-Heisenberg-Institut) \\
F\"ohringer Ring 6, 80805 M\"unchen, Germany}

\begin{abstract}
We discuss the effects of neutrino mixing and the neutrino mass hierarchy when considering the capture of the cosmic neutrino background (CNB) on radioactive nuclei. The implications of mixing and hierarchy at future generations of tritium decay experiments are considered. We find that the CNB should be detectable at these experiments provided that the resolution for the kinetic energy of the outgoing electron can be pushed to a few 0.01~eV for the scenario with inverted neutrino mass hierarchy, about an order of magnitude better than that of the upcoming KATRIN experiment. Another order of magnitude improvement is needed in the case of normal neutrino mass hierarchy. We also note that mixing effects generally make the prospects for CNB detection worse due to an increased maximum energy of the normal beta decay background.
\end{abstract}


\preprint{MPP-2008-23}

\maketitle

\section{Introduction}

Recently, there have been a number of papers \cite{Cocco:2007za,Lazauskas:2007da} on the possible detection of the $T_\nu \simeq 1.9$~K cosmic neutrino background (CNB) (see, \eg, \Ref~\cite{Weinberg}, for a recent review of neutrinos in cosmology, see \Ref~\cite{Dolgov:2002wy}) in beta decay experiments initially designed to measure the neutrino mass through the kinematics of the decay. Although a fairly old idea \cite{Irvine:1983nr}, this prospect now seems more feasible as the energy resolution of the experiments are being pushed down to sub-eV scales (\eg, by the KATRIN experiment~\cite{Osipowicz:2001sq}). Among other things, \Refs~\cite{Cocco:2007za,Lazauskas:2007da} discuss how the signal-to-background ratio for such a CNB search depends on the energy resolution and neutrino mass. Specifically, it was noted in \Ref~\cite{Lazauskas:2007da} that an energy resolution of $\Delta \leq 0.5$~eV would be required in order to detect the CNB if the neutrino mass is on the sub-eV scale, independent of gravitational clustering of cosmic neutrinos. For a summary of other methods which have been discussed for measuring the CNB, see \Ref~\cite{Gelmini:2004hg}.

In this text, we will discuss the effects of neutrino mixing and the neutrino mass hierarchy on the above mentioned CNB capture. Clearly, if neutrinos are degenerate in mass, then the effects of mixing and hierarchy will be small. However, if neutrinos have hierarchical masses, then also the neutrino mixing will affect the prospects of CNB detection. Unfortunately, hierarchical masses also require the neutrinos to be very light, making the experimental detection extremely difficult and demanding. However, there is a hierarchy dependent limit on the resolution for when the CNB should be detected due to the known mass squared differences.

\section{Neutrino masses and mixing}

Since the Super-Kamiokande results in 1998~\cite{Fukuda:1998mi}, different experiments have been adding more and more evidence to the existence of neutrino oscillations and today the neutrino oscillation parameters are fairly well-constrained~\cite{Maltoni:2004ei}. The basic principle behind neutrino oscillations is based on the neutrino weak interaction eigenstates $\ket{\nu_\alpha}$ being different from the neutrino mass eigenstates $\ket{\nu_i}$. In analogy to the quark mixing, the weak interaction and mass eigenstates of the neutrinos are related as
\begin{equation}
\ket{\nu_\alpha} = \sum_i U^*_{\alpha i} \ket{\nu_i},
\end{equation}
where $U_{\alpha i}$ are the elements of the leptonic analogue of the Cabibbo--Kobayashi--Maskawa matrix. Furthermore, neutrino oscillation experiments imply that neutrinos are massive, although they can only probe the neutrino mass squared differences $\Delta m_{ij}^2 = m_i^2-m_j^2$, where $m_i$ is the mass of $\nu_i$, and not the absolute neutrino mass scale. 
While many of the neutrino oscillation parameters are quite well constrained, the questions whether $\theta_{13}$ is zero and whether $\Delta m_{31}^2$ is positive (normal neutrino mass hierarchy) or negative (inverted neutrino mass hierarchy) remain. In the remainder of this text, we will use $\sin^2(\theta_{12}) = 0.3$, $|\Delta m_{31}^2| = 2.2\cdot 10^{-3}$~eV$^2$, and $\Delta m_{21}^2 = 7.6\cdot 10^{-5}$~eV$^2$. These values are the best-fit values of \Ref~\cite{Maltoni:2004ei} except for the value of $\Delta m_{21}^2$, which has been inspired by the recent publication of precision data by KamLAND~\cite{Collaboration:2008ee}. As will be seen, the value of $\theta_{23}$ and any phases (either Dirac or Majorana) are irrelevant to the process in question and the value of $\theta_{13}$ as well as the neutrino mass hierarchy will be varied throughout.

In the following, we will denote the mass of a general neutrino by $m_\nu$ and the mass of the lightest neutrino by $m_0$. With this notation, we have
\begin{equation}
m_1 = m_0, \quad m_2 = \sqrt{m_0^2 + \Delta m_{21}^2}, \quad m_3 = \sqrt{m_0^2+|\Delta m_{31}^2|}
\end{equation}
in the case of normal neutrino mass hierarchy and
\begin{equation}
m_1 = \sqrt{m_0^2+|\Delta m_{31}^2|}, \quad m_2 = \sqrt{m_0^2+|\Delta m_{31}^2|+\Delta m_{21}^2}, \quad m_3 = m_0
\end{equation}
in the case of inverted neutrino mass hierarchy.

\section{Physics of the neutrino capture}

The physics for the neutrino capture $\nu_e + X \rightarrow e^- + Y$ by the beta decaying nucleus $X$ (decaying to $Y$ via the process $X \rightarrow \bar\nu_e + e^- + Y$) is very straightforward and the capture rate can easily be found as \cite{Lazauskas:2007da}
\begin{equation}
\label{eq:eventrate}
N_{\rm CNB} \simeq 6.5 \rho_{\rm c}~{\rm yr}^{-1}{\rm MCi}^{-1},
\end{equation} 
for a sample of tritium, where $\rho_c = n_\nu/\langle n_\nu\rangle$ is the ratio of the relic neutrino density at Earth and the mean relic neutrino density in the Universe (we expect to have $\rho_{\rm c} \geq 1$ due to gravitational clustering of relic neutrinos). The average density of a single neutrino state can be computed to be $\langle n_\nu\rangle \simeq 55$~cm$^{-3}$.
While interaction and oscillation effects in the early Universe can slightly modify this number~\cite{Mangano:2005cc}, the corrections are not of practical importance for the present study.
Thus, for a 1~MCi source of tritium (roughly 100~g) and depending on the  clustering and running time of the experiment, we could have a significant number of events. The main problem is to separate the CNB events from the background of the usual beta decay process. In order to achieve this, the energy resolution for the experiment needs to be $\Delta \lesssim 0.5 m_\nu$ in order to detect the CNB for sub-eV neutrinos \cite{Lazauskas:2007da} ($\Delta$ being the full-width at half-maximum of a gaussian energy resolution for the outgoing electrons).

Equation~(\ref{eq:eventrate}) is computed for a single neutrino and will result in electrons of energy $Q+m_\nu$, where $Q$ is the energy release in the usual beta decay process assuming the neutrino to have zero mass. When considering mixed neutrinos with each neutrino mass eigenstate having a relic density of $n_{\nu_i} = \rho_{i,\rm c} \langle n_\nu\rangle$, the capture rate for each mass eigenstate is
\begin{equation}
N_{i,\rm CNB} = 6.5 \rho_{i,\rm c} |U_{ei}|^2~{\rm yr}^{-1}{\rm MCi}^{-1},
\end{equation}
where $|U_{ei}|^2$ is the electron neutrino content of $\nu_i$, each giving monoenergetic electrons with kinetic energy $Q+m_i$. With the knowledge from neutrino oscillation experiments, we have
\begin{equation}
|U_{e1}|^2 = c_{13}^2 c_{12}^2 \simeq 0.7 c_{13}^2, \quad
|U_{e2}|^2 = c_{13}^2 s_{12}^2 \simeq 0.3 c_{13}^2, \quad
|U_{e3}|^2 = s_{13}^2,
\end{equation}
where $s_{ij} = \sin(\theta_{ij})$ and $c_{ij} = \cos(\theta_{ij})$. Note that the lepton mixing angle $\theta_{23}$ is not present, since it does not affect the mixing of the electron neutrino. In principle, the clustering ratios $\rho_{i,\rm c}$ could be different for different mass eigenstates. As we will mainly be interested in the worst-case scenario with hierarchical neutrino masses of the order of $0.05$~eV or smaller, we will assume that all clustering coefficients $\rho_{i,\rm c} = 1$ (\cf~\Ref~\cite{Ringwald:2004np}). As mentioned in \Ref~\cite{Lazauskas:2007da}, the clustering ratios do not significantly affect the resolution needed in order to obtain a good signal-to-noise ratio. However, any clustering will, of course, lead to an enhanced signal rate.

For the background involving the standard tritium beta decay, we note that its magnitude is trivially related to the signal, since both the background rate and the signal rate involve the very same matrix element. The background spectrum is given by \cite{McKellar:1980cn}
\begin{equation}
\dd{N_\beta}{T_e} = \sum_i |U_{ei}|^2 \dd{N_{\beta,i}}{T_e},
\end{equation}
where $\tdd{N_{\beta,i}}{T_e}$ is the spectrum for the beta decay assuming a single unmixed neutrino of mass $m_\nu = m_i$ and $T_e$ is the kinetic energy of the outgoing electron. Because of the kinematics of this process, $\tdd{N_{\beta,i}}{T_e}$ is zero above $T_e = Q-m_i$. However, due to the overwhelming number of events, a large number of events may appear even at the energies $T_e = Q+m_i$ because of the finite energy resolution of any experiment.

\section{Mixing effects on the spectrum}

For degenerate neutrino masses (\ie, $m_0 \gg \sqrt{|\Delta m_{31}^2|}$) it is trivial to find that the effects of neutrino mixing on the capture of CNB on beta decaying nuclei are negligible. Since $\sum_i |U_{ei}|^2 = 1$ (assuming unitarity of $U$) and the neutrino masses are nearly equal (and thus also the clustering of different mass eigenstates), the signal will be the same as for a single neutrino with no mixing as long as the experiment is not sensitive enough to detect the very small neutrino mass differences. For example, in the case of $m_0 \simeq 1$~eV, the neutrino mass difference would be $|m_3-m_1| \simeq |\Delta m_{31}^2|/(2m_0) \simeq 1.1\cdot 10^{-3}$~eV, meaning that a resolution better than $\Delta \simeq 0.5$~meV would be needed in order to tell the peaks from the different mass eigenstates apart.

\subsection{The case of small $\boldsymbol{\theta_{13}}$}

\begin{figure}
\begin{center}
\includegraphics[width=0.7\textwidth]{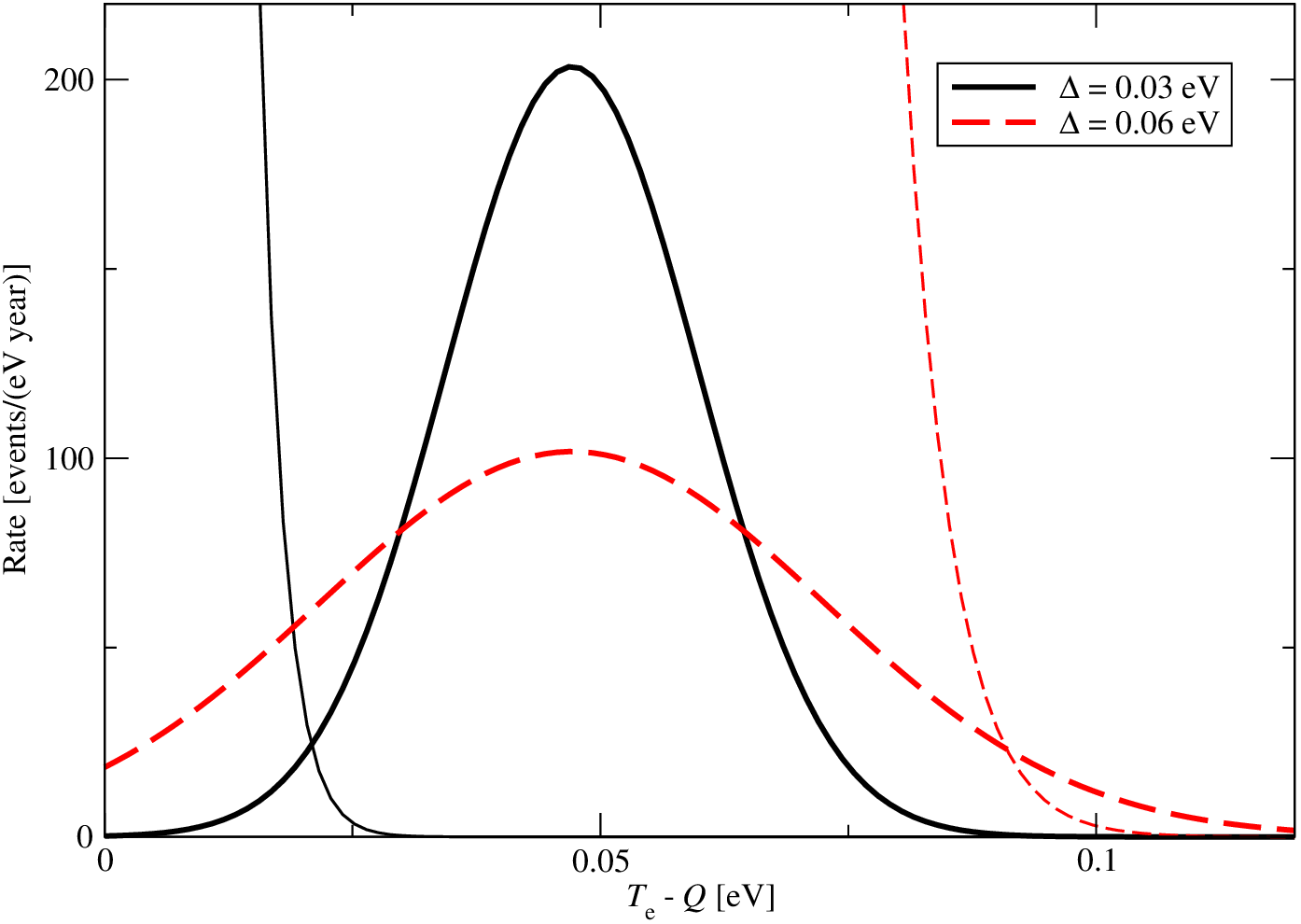}
\caption{Event rates as functions of the kinetic energy $T_e$ of the electron for inverted hierarchy, $\theta_{13} = 0$, and gaussian detector resolution $\Delta$. The bold (thin) curves correspond to the CNB signal (beta decay background) and we have assumed a negligible gravitational clustering.}
\label{fig:theta0inv}
\end{center}
\end{figure}

With the above in mind, let us focus on the case when the neutrino masses are hierarchical ($m_0 \ll \sqrt{\Delta m_{21}^2}$), for definiteness, we use $m_0 = 0$. We first assume that the lepton mixing angle $\theta_{13} = 0$ and later consider the implications of non-zero $\theta_{13}$. With $\theta_{13} = 0$, we have a situation where only two of the neutrino mass eigenstates are involved in the beta decay and the CNB capture. The prospects of CNB detection is now ultimately dependent on the neutrino mass hierarchy. For an inverted hierarchy, the mass eigenstates involved have masses of $m_{1,2} \simeq \sqrt{|\Delta m_{31}^2|} \simeq 0.047$~eV with a mass difference of $m_2-m_1 \simeq \Delta m_{21}^2/(2m_{1,2}) \simeq 8\cdot 10^{-4}$~eV. As in the case of degenerate masses, experiments with enough accuracy to separate the two mass eigenstates seem extremely unlikely at the present time. However, an experiment with enough accuracy to detect the CNB with a reasonable signal-to-background ratio would have to have a resolution of $\Delta \simeq 0.03$~eV (see \fig~\ref{fig:theta0inv}).
It should be noted that, while this resolution may also seem extreme, it represents a resolution at which the CNB could be detected regardless of the absolute neutrino mass scale (also note that the resolution for which the signal-to-noise ratio becomes of order unity is not very dependent on the gravitational clustering \cite{Lazauskas:2007da}).

\begin{figure}
\begin{center}
\includegraphics[width=0.7\textwidth]{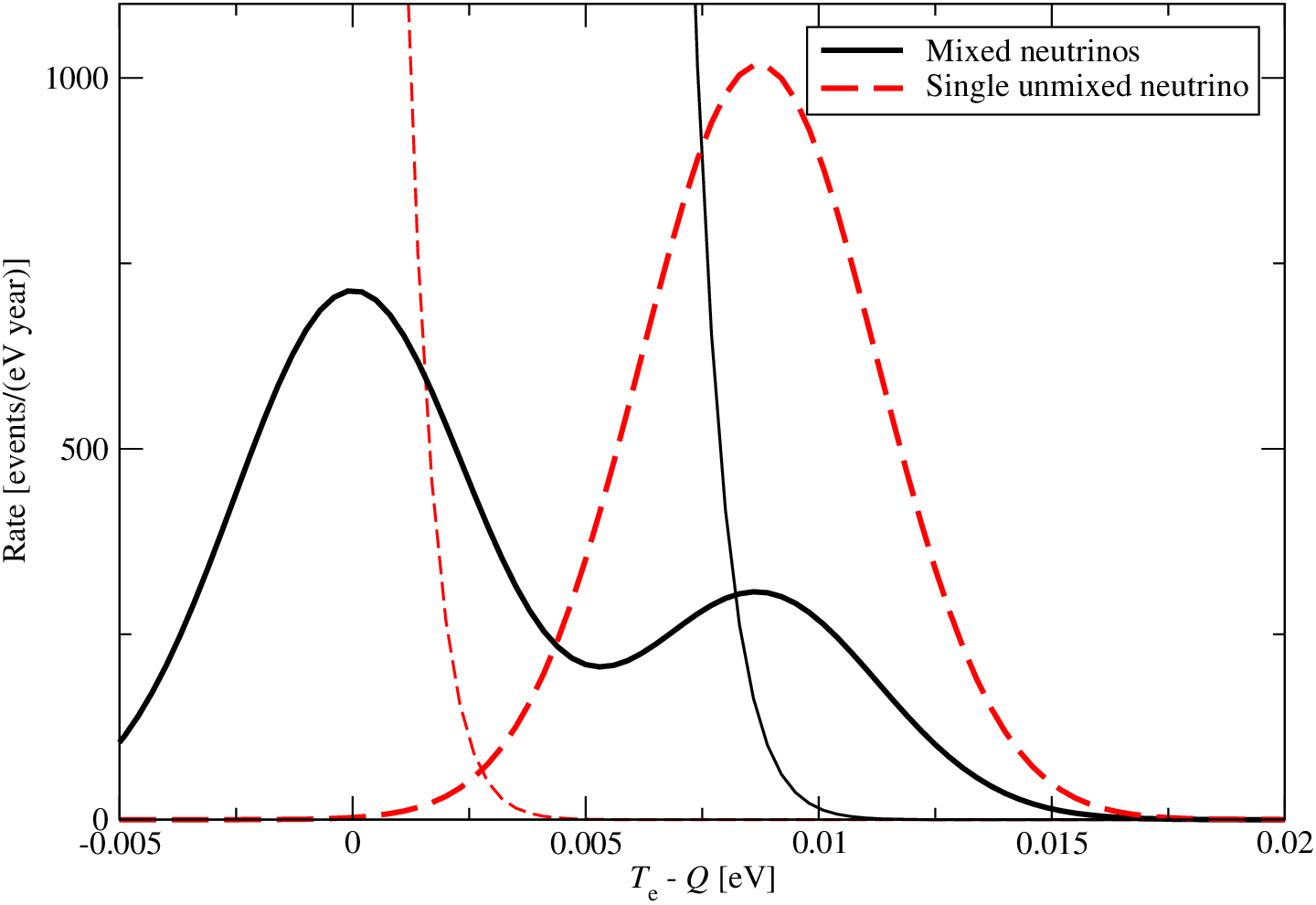}
\caption{Event rates as functions of the kinetic energy $T_e$ of the electron for the normal hierarchy, $\theta_{13} = 0$, and a gaussian detector resolution of $\Delta = 0.008$~eV. The bold (thin) curves correspond to the CNB signal (beta decay background) and we have assumed a negligible gravitational clustering.}
\label{fig:theta0norm}
\end{center}
\end{figure}

For the case of normal neutrino mass hierarchy, the situation becomes worse. In this scenario, the neutrino masses involved are $m_1 = 0$ and $m_2 = \sqrt{\Delta m_{21}^2} \simeq 0.01$~eV, respectively. Since one of the mass eigenstates has zero mass, the background spectrum will reach as high as the total release energy $Q$, basically obstructing the possibility of detecting the peak for $m_1$. Assuming no clustering, the signal-to-noise ratio would be of order one at $\Delta \simeq 10^{-4}$~eV, using the relation
\begin{equation}
\frac{N_{\rm CNB}}{N_\beta} \simeq 6\pi^2 \frac{n_\nu}{\Delta^3} \simeq 2.5\cdot 10^{-11} \rho_{\rm c} \left(\frac{1~\rm eV}{\Delta}\right)^3
\label{eq:ratioapprox}
\end{equation}
for the number of events within $\Delta$ from the end-point energy. The resolution needed to detect the CNB peak for the second mass eigenstate would also need to be better than in the case with only one unmixed neutrino of the same mass (see \fig~\ref{fig:theta0norm}). This is due to the fact that in the latter case, the maximal energy of the background electrons from the standard beta decay is $Q-m$ rather than $Q$.
Furthermore, the event rate in the case of mixed neutrinos is less by a factor of $s_{12}^2 \simeq 0.3$, since the contribution from the first mass eigenstate is not detectable. In order to obtain a good signal-to-background ratio for the peak corresponding to the second mass eigenstate, an energy resolution of a few meV seems to be necessary.

\subsection{The case of large $\boldsymbol{\theta_{13}}$}

If the lepton mixing angle $\theta_{13}$ turns out to have a relatively large value (we keep in mind that the upper bound on $\theta_{13}$ from the CHOOZ experiment~\cite{Apollonio:1999ae} is $s_{13}^2 \leq 0.047$ at 3$\sigma$, corresponding to $\theta_{13} \leq 12^\circ$), then this will affect the conclusions from the previous subsection. This will imply that at least one neutrino of mass $m_i \simeq \sqrt{|\Delta m_{31}^2|} \simeq 0.047$~eV will always be involved in the processes under scrutiny. Thus, an energy resolution comparable to this value might be able to produce a reasonable signal-to-background ratio regardless of the neutrino hierarchy. Clearly, this is positive, since it would imply that there is a lower bound on the resolution where one could expect to detect the CNB assuming that the event rate can be made high enough. However, there is also the drawback of always involving a neutrino with zero mass, resulting in the extension of the background spectrum up to $T_e = Q$, and thus, requiring better energy resolution than if only involving one massive neutrino.

\begin{figure}
\begin{center}
\includegraphics[width=0.7\textwidth]{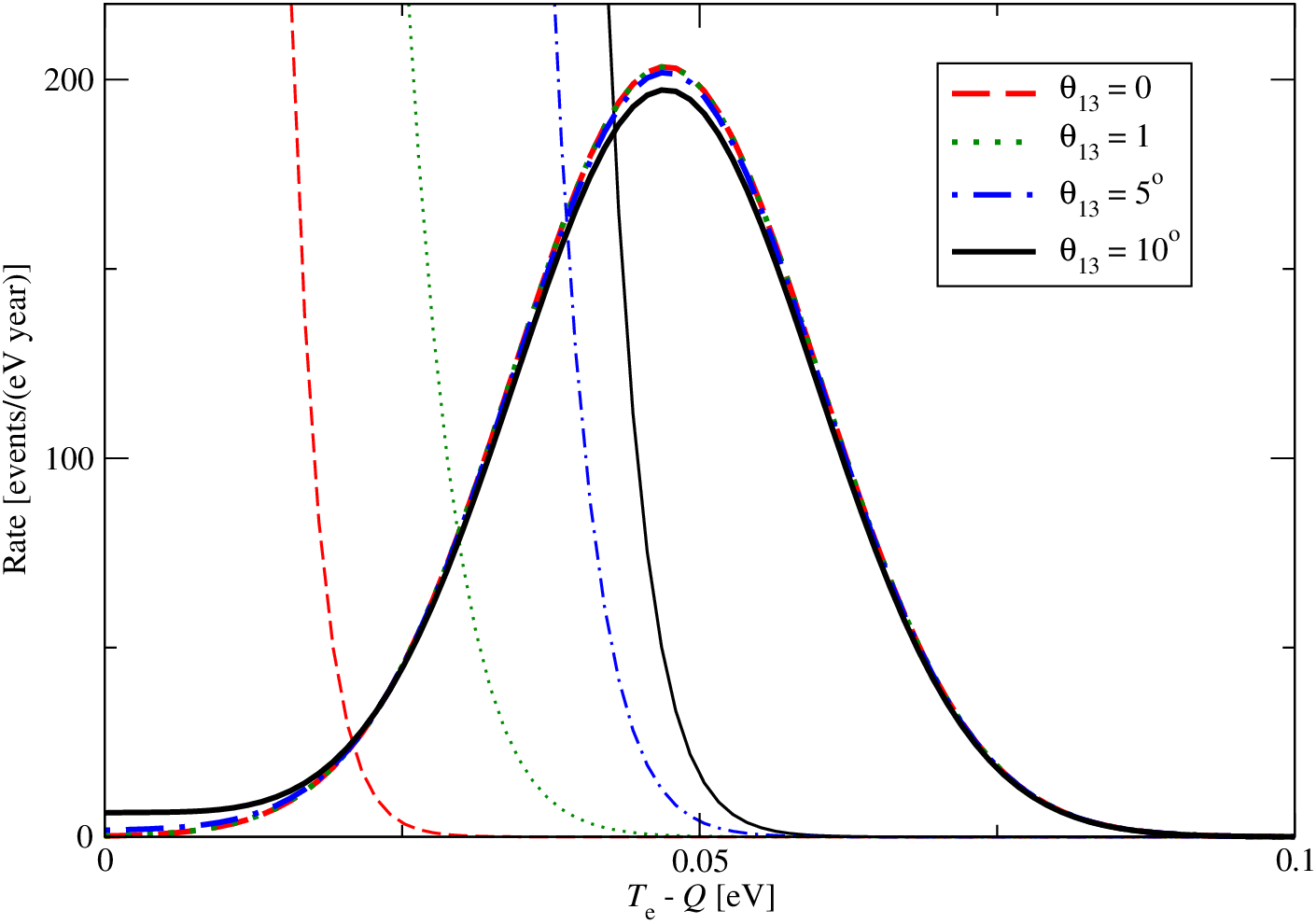}
\caption{Event rates as functions of the kinetic energy $T_e$ of the electron for inverted hierarchy, different $\theta_{13}$, and gaussian detector resolution $\Delta = 0.03$~eV. The bold (thin) curves correspond to the CNB signal (beta decay background) and we have assumed a negligible gravitational clustering (\cf, \fig~\ref{fig:theta0inv}).}
\label{fig:theta10inv}
\end{center}
\end{figure}

In \fig~\ref{fig:theta10inv}, we show the inverted hierarchy result for $\theta_{13} = 0, 1^\circ, 5^\circ$ and $10^\circ$. As can be seen from this figure, when $\theta_{13} \neq 0$, the signal itself is only slightly suppressed due to the small mixing of the third mass eigenstate. However, the background from beta decays involving the neutrino mass eigenstate $\nu_3$ is still sufficient to produce a significant increase in the background, resulting in the need for slightly better energy resolution than if $\theta_{13} = 0$.
Since the background related to the third mass eigenstate basically grows exponentially with decreasing $T_e$, even a small $\theta_{13}$ is enough to increase the background significantly.

\begin{figure}
\begin{center}
\includegraphics[width=0.7\textwidth]{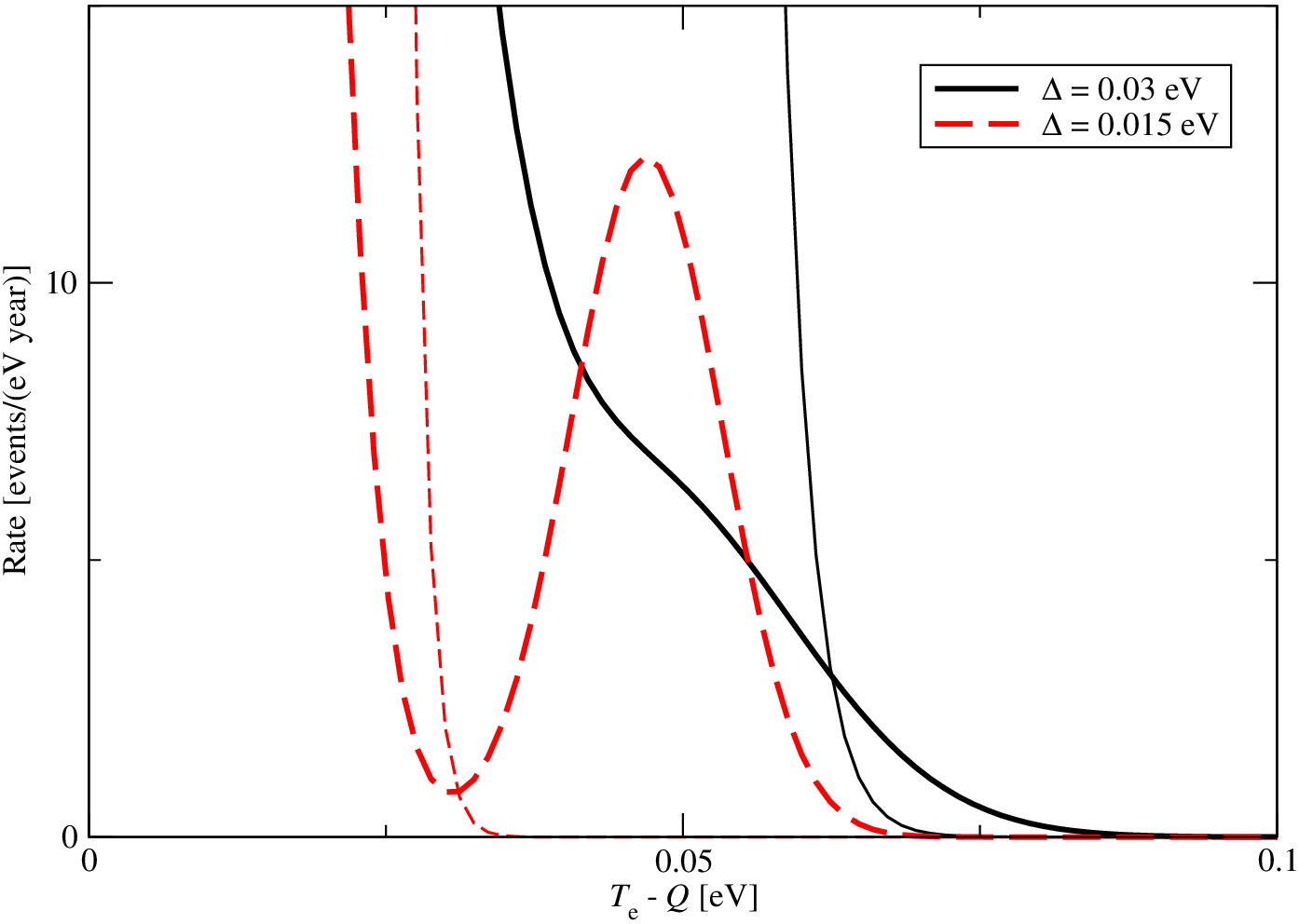}
\caption{Event rates as functions of the kinetic energy $T_e$ of the electron for inverted hierarchy, $\theta_{13} = 10^\circ$, and gaussian detector resolution $\Delta$. The bold (thin) curves correspond to the CNB signal (beta decay background) and we have assumed a negligible gravitational clustering.}
\label{fig:theta10norm}
\end{center}
\end{figure}

The case of $\theta_{13} = 10^\circ$ and normal mass hierarchy is shown in \fig~\ref{fig:theta10norm}.
As can be observed, the background for the energy resolution $\Delta = 0.03$~eV reaches slightly higher than in the case of the inverted mass hierarchy. This is simply due to the low-mass neutrinos having a larger mixing with the electron neutrino, resulting in higher maximum energy for the background beta decays with the larger rates. In addition, the event rate corresponding to the third mass eigenstate is significantly lower that that of the first and second mass eigenstates in the case of inverted neutrino mass hierarchy, simply due to the small value of $\theta_{13}$. Thus, a large value of $s_{13}^2 \rho_{3,\rm c}$ would be needed in order to have a sufficient event rate. Supposing that this could happen, the peak could be resolved with an energy resolution of $\Delta \simeq 0.015$~eV, which is about a factor of four less requiring than the resolution needed to resolve the second mass eigenstate in the normal hierarchy when $\theta_{13} = 0$. The requirements for resolving the second mass eigenstate is practically unchanged as the background is more or less unaffected by $\theta_{13}$ in the normal neutrino mass hierarchy.

\section{A note on the case of extreme gravitational clustering}

It has been speculated that an extreme clustering of $\rho_{\rm c} = 10^{13}$ for $m_\nu = 0.5$~eV neutrinos could be responsible for the knee in the cosmic ray spectrum due to the threshold of the $p + \bar \nu_e \rightarrow n + e^+$ reaction \cite{Wigmans:2002rb,Hwang:2005dq}. In this scenario, the rate of electrons originating in beta decay and the rate of electrons originating in the capture of CNB neutrinos in the last energy bin can again be approximated by \eq~(\ref{eq:ratioapprox}). However, with this extreme clustering the ratio of events due to CNB interactions and events due to beta decay is given by
\begin{equation}
\frac{N_{\rm CNB}}{N_\beta} \simeq 6\pi^2 \frac{n_\nu}{\Delta^3} \simeq 250 \left(\frac{1~\rm eV}{\Delta}\right)^3.
\end{equation}
With an energy resolution of about 5~eV, comparable to that of the Troitsk experiment~\cite{Lobashev:2001uu}, this would imply a ratio of $N_{\rm CNB}/N_\beta \simeq 2$. Thus, the effect of CNB neutrinos on the electron spectrum would definitely be non-negligible in this scenario and, if not taken into account, would lead to an over-estimate in the endpoint energy, clearly consistent with the negative best-fit value of $m_\nu^2$ at Troitsk ($m_\nu^2 = -1.0\pm 5.1$~eV$^2$~\cite{Lobashev:2001uu}). If this extreme gravitational clustering of $m_\nu = 0.5$~eV background neutrinos is present at the Earth, then the discovery of the CNB should be just around the corner with the upcoming results of the KATRIN experiment \cite{Osipowicz:2001sq}, sensitive to neutrino masses of about $m_\nu \simeq 0.2$~eV. However, if it is not detected by KATRIN, then this would seem to invalidate (at least locally) the assumption of extremely clustered background neutrinos. Since the ratio is proportional to $\Delta^{-3}$, KATRIN would be sensitive to CNB neutrinos with a clustering of $\rho_c \gtrsim 10^8$ regardless of the neutrino mass.

\section{Summary and discussion}

We have studied the implications of lepton mixing and the neutrino mass hierarchy on the possible detection of the CNB in tritium beta decay experiments and how the energy resolution for the outgoing electrons affect the signal-to-background ratio, where the background consists of electrons from the normal beta decay. We have found that even if neutrinos are strongly hierarchical with the lightest neutrino mass being zero, the CNB should be detectable with a reasonable rate in an experiment involving about 100~g of tritium as long as an energy resolution of a few 0.01~eV can be achieved and the neutrino mass hierarchy is inverted. If we have a normal neutrino mass hierarchy, then the situation is worse unless the event rate corresponding to the capture of $\nu_3$ can be increased (\eg, by a larger sample or a very large value of $s_{13}^2\rho_{3,\rm c}$). Thus, reaching this accuracy without observing the CNB signal with the rate expected from the inverted hierarchy would imply a normal neutrino mass hierarchy. 
However, it is impossible to distinguish the hierarchies if a single peak at an energy $T_e \geq Q+\sqrt{\Delta m_{31}^2}$ is found. This is due to the fact that the $\nu_{1,2}$ peak can be in this region in both hierarchies, depending on $m_0$.
Furthermore, the observation of a CNB signal would give us experimental insight into the amount of clustering. It should however be noted that the difference in the signal strength for the normal and inverted mass hierarchies would imply different amounts of clustering.

In the event of normal mass hierarchy, a detector resolution of a few meV would be needed in order to have a reasonable signal-to-background ratio for the peak corresponding to the second neutrino mass eigenstate. If $\theta_{13}$ is shown to be relatively large, then we would still need enough energy resolution to resolve both the peak for $m_2$ and $m_3$ in the case of the normal hierarchy in order to tell which hierarchy is correct. This is due to the fact that the $m_{1,2}$ peak in the case of inverted hierarchy could also result from the $m_3$ peak in the normal hierarchy if the gravitational clustering is large. In this case, the relative strength of the peaks would tell which peak belongs to which mass eigenstate, and thus, discriminate between the neutrino mass hierarchies.

We have also noted that in the case of extreme gravitational clustering of the order of $10^{13}$, the CNB should have had a reasonably large effect already at the Troitsk experiment. In this case, the detection of the CNB should be right around the corner with the advent of the results of the KATRIN experiment.

A final remark; at the energy resolutions discussed here, the nuclear recoil energy of $E_{\rm recoil} \simeq 2Q^2/M_{{}^3{\rm He}} \simeq 0.25$~eV can no longer be neglected. However, this can be readily taken into account in the analysis of any experiment. At the even lower energy resolutions discussed, finite temperature effects in the tritium source may also play a role. As comparison, temperature effects in the KATRIN experiment of the order of $T = 27~{\rm K} \sim 2$~meV are negligible compared to the energy resolution. The lower limit for the temperature broadening effect in the CNB capture is set by the CNB temperature of $T_{\rm CNB} = 1.9~{\rm K} \sim 0.16$~meV.

\subsection*{Acknowledgments}

The author is grateful to Tommy Ohlsson and Alessandro Mirizzi for reading different drafts of this paper and making useful comments. This work was supported by the Swedish Research Council (Vetenskapsr\aa{}det) through Contract No.~623-2007-8066.


\begin{thebibliography}{17}
\expandafter\ifx\csname natexlab\endcsname\relax\def\natexlab#1{#1}\fi
\expandafter\ifx\csname bibnamefont\endcsname\relax
  \def\bibnamefont#1{#1}\fi
\expandafter\ifx\csname bibfnamefont\endcsname\relax
  \def\bibfnamefont#1{#1}\fi
\expandafter\ifx\csname citenamefont\endcsname\relax
  \def\citenamefont#1{#1}\fi
\expandafter\ifx\csname url\endcsname\relax
  \def\url#1{\texttt{#1}}\fi
\expandafter\ifx\csname urlprefix\endcsname\relax\def\urlprefix{URL }\fi
\providecommand{\bibinfo}[2]{#2}
\providecommand{\eprint}[2][]{\url{#2}}

\bibitem[{\citenamefont{Cocco et~al.}(2007)\citenamefont{Cocco, Mangano, and
  Messina}}]{Cocco:2007za}
\bibinfo{author}{\bibfnamefont{A.~G.} \bibnamefont{Cocco}},
  \bibinfo{author}{\bibfnamefont{G.}~\bibnamefont{Mangano}}, \bibnamefont{and}
  \bibinfo{author}{\bibfnamefont{M.}~\bibnamefont{Messina}},
  \bibinfo{journal}{JCAP} \textbf{\bibinfo{volume}{0706}}, \bibinfo{pages}{015}
  (\bibinfo{year}{2007}), \eprint{hep-ph/0703075}.

\bibitem[{\citenamefont{Lazauskas et~al.}(2008)\citenamefont{Lazauskas, Vogel,
  and Volpe}}]{Lazauskas:2007da}
\bibinfo{author}{\bibfnamefont{R.}~\bibnamefont{Lazauskas}},
  \bibinfo{author}{\bibfnamefont{P.}~\bibnamefont{Vogel}}, \bibnamefont{and}
  \bibinfo{author}{\bibfnamefont{C.}~\bibnamefont{Volpe}}, \bibinfo{journal}{J.
  Phys.} \textbf{\bibinfo{volume}{G35}}, \bibinfo{pages}{025001}
  (\bibinfo{year}{2008}), \eprint{arXiv:0710.5312 [astro-ph]}.

\bibitem[{\citenamefont{Weinberg}(1972)}]{Weinberg}
\bibinfo{author}{\bibfnamefont{S.}~\bibnamefont{Weinberg}},
  \emph{\bibinfo{title}{Gravitation and cosmology}} (\bibinfo{year}{1972}).

\bibitem[{\citenamefont{Dolgov}(2002)}]{Dolgov:2002wy}
\bibinfo{author}{\bibfnamefont{A.~D.} \bibnamefont{Dolgov}},
  \bibinfo{journal}{Phys. Rept.} \textbf{\bibinfo{volume}{370}},
  \bibinfo{pages}{333} (\bibinfo{year}{2002}), \eprint{hep-ph/0202122}.

\bibitem[{\citenamefont{Irvine and Humphreys}(1983)}]{Irvine:1983nr}
\bibinfo{author}{\bibfnamefont{J.~M.} \bibnamefont{Irvine}} \bibnamefont{and}
  \bibinfo{author}{\bibfnamefont{R.}~\bibnamefont{Humphreys}},
  \bibinfo{journal}{J. Phys.} \textbf{\bibinfo{volume}{G9}},
  \bibinfo{pages}{847} (\bibinfo{year}{1983}).

\bibitem[{\citenamefont{Osipowicz et~al.}(2001)}]{Osipowicz:2001sq}
\bibinfo{author}{\bibfnamefont{A.}~\bibnamefont{Osipowicz}}
  \bibnamefont{et~al.} (\bibinfo{collaboration}{KATRIN})
  (\bibinfo{year}{2001}), \eprint{hep-ex/0109033}.

\bibitem[{\citenamefont{Gelmini}(2005)}]{Gelmini:2004hg}
\bibinfo{author}{\bibfnamefont{G.~B.} \bibnamefont{Gelmini}},
  \bibinfo{journal}{Phys. Scripta} \textbf{\bibinfo{volume}{T121}},
  \bibinfo{pages}{131} (\bibinfo{year}{2005}), \eprint{hep-ph/0412305}.

\bibitem[{\citenamefont{Fukuda et~al.}(1998)}]{Fukuda:1998mi}
\bibinfo{author}{\bibfnamefont{Y.}~\bibnamefont{Fukuda}} \bibnamefont{et~al.}
  (\bibinfo{collaboration}{Super-Kamiokande}), \bibinfo{journal}{Phys. Rev.
  Lett.} \textbf{\bibinfo{volume}{81}}, \bibinfo{pages}{1562}
  (\bibinfo{year}{1998}), \eprint{hep-ex/9807003}.

\bibitem[{\citenamefont{Maltoni et~al.}(2004)\citenamefont{Maltoni, Schwetz,
  Tortola, and Valle}}]{Maltoni:2004ei}
\bibinfo{author}{\bibfnamefont{M.}~\bibnamefont{Maltoni}},
  \bibinfo{author}{\bibfnamefont{T.}~\bibnamefont{Schwetz}},
  \bibinfo{author}{\bibfnamefont{M.~A.} \bibnamefont{Tortola}},
  \bibnamefont{and} \bibinfo{author}{\bibfnamefont{J.~W.~F.}
  \bibnamefont{Valle}}, \bibinfo{journal}{New J. Phys.}
  \textbf{\bibinfo{volume}{6}}, \bibinfo{pages}{122} (\bibinfo{year}{2004}),
  \bibinfo{note}{results in eprint updated since publication, see references
  therein for the experimental results leading to the parameter constraints},
  \eprint{hep-ph/0405172v6}.

\bibitem[{\citenamefont{Abe et~al.}(2008)}]{Collaboration:2008ee}
\bibinfo{author}{\bibfnamefont{S.}~\bibnamefont{Abe}} \bibnamefont{et~al.}
  (\bibinfo{collaboration}{KamLAND collaboration}) (\bibinfo{year}{2008}),
  \eprint{arXiv:0801.4589 [hep-ex]}.

\bibitem[{\citenamefont{Mangano et~al.}(2005)}]{Mangano:2005cc}
\bibinfo{author}{\bibfnamefont{G.}~\bibnamefont{Mangano}} \bibnamefont{et~al.},
  \bibinfo{journal}{Nucl. Phys.} \textbf{\bibinfo{volume}{B729}},
  \bibinfo{pages}{221} (\bibinfo{year}{2005}), \eprint{hep-ph/0506164}.

\bibitem[{\citenamefont{Ringwald and Wong}(2004)}]{Ringwald:2004np}
\bibinfo{author}{\bibfnamefont{A.}~\bibnamefont{Ringwald}} \bibnamefont{and}
  \bibinfo{author}{\bibfnamefont{Y.~Y.~Y.} \bibnamefont{Wong}},
  \bibinfo{journal}{JCAP} \textbf{\bibinfo{volume}{0412}}, \bibinfo{pages}{005}
  (\bibinfo{year}{2004}), \eprint{hep-ph/0408241}.

\bibitem[{\citenamefont{McKellar}(1980)}]{McKellar:1980cn}
\bibinfo{author}{\bibfnamefont{B.~H.~J.} \bibnamefont{McKellar}},
  \bibinfo{journal}{Phys. Lett.} \textbf{\bibinfo{volume}{B97}},
  \bibinfo{pages}{93} (\bibinfo{year}{1980}).

\bibitem[{\citenamefont{Apollonio et~al.}(1999)}]{Apollonio:1999ae}
\bibinfo{author}{\bibfnamefont{M.}~\bibnamefont{Apollonio}}
  \bibnamefont{et~al.} (\bibinfo{collaboration}{CHOOZ}),
  \bibinfo{journal}{Phys. Lett.} \textbf{\bibinfo{volume}{B466}},
  \bibinfo{pages}{415} (\bibinfo{year}{1999}), \eprint{hep-ex/9907037}.

\bibitem[{\citenamefont{Wigmans}(2003)}]{Wigmans:2002rb}
\bibinfo{author}{\bibfnamefont{R.}~\bibnamefont{Wigmans}},
  \bibinfo{journal}{Astropart. Phys.} \textbf{\bibinfo{volume}{19}},
  \bibinfo{pages}{379} (\bibinfo{year}{2003}), \eprint{astro-ph/0205360}.

\bibitem[{\citenamefont{Hwang and Ma}(2005)}]{Hwang:2005dq}
\bibinfo{author}{\bibfnamefont{W.-Y.~P.} \bibnamefont{Hwang}} \bibnamefont{and}
  \bibinfo{author}{\bibfnamefont{B.-Q.} \bibnamefont{Ma}},
  \bibinfo{journal}{New J. Phys.} \textbf{\bibinfo{volume}{7}},
  \bibinfo{pages}{41} (\bibinfo{year}{2005}), \eprint{astro-ph/0502377}.

\bibitem[{\citenamefont{Lobashev et~al.}(2001)}]{Lobashev:2001uu}
\bibinfo{author}{\bibfnamefont{V.~M.} \bibnamefont{Lobashev}}
  \bibnamefont{et~al.}, \bibinfo{journal}{Nucl. Phys. Proc. Suppl.}
  \textbf{\bibinfo{volume}{91}}, \bibinfo{pages}{280} (\bibinfo{year}{2001}).

\end{thebibliography}
\end{document}